  \documentclass[prr,twocolumn,nopacs,superscriptaddres,amsmath]{revtex4}
\usepackage{dcolumn}
\usepackage{bm}
\usepackage{graphicx}
\usepackage{color}
\usepackage{subfigure}
\usepackage{hyperref}
\usepackage{latexsym}
\usepackage{amsthm}
\usepackage{amssymb}
\usepackage{array}
\usepackage{braket}

\DeclareGraphicsExtensions{.jpg,.pdf, .mps, .png, .eps, .ps, .EPS,.gif}

\begin{document}
\def\be{\begin{equation}}
\def\ee{\end{equation}}
	
\def\bc{\begin{center}}
\def\ec{\end{center}}
\def\bea{\begin{eqnarray}}
\def\eea{\end{eqnarray}}
\newcommand{\avg}[1]{\langle{#1}\rangle}
\newcommand{\Avg}[1]{\left\langle{#1}\right\rangle}

\def\ie{\textit{i.e.}}
\def\etal{\textit{et al.}}
\def\m{\vec{m}}
\def\G{\mathcal{G}}

\newcommand{\giacomo}[1]{{\bf\color{cyan}#1}}
\newcommand{\gin}[1]{{\bf\color{green}#1}}
\newcommand{\alex}[1]{{\bf\color{red}#1}}
\newcommand{\hanlin}[1]{{\bf\color{blue}#1}}

\title{{A message-passing approach to epidemic tracing and mitigation with apps}}

\author{Ginestra Bianconi}
\affiliation{School of Mathematical Sciences, Queen 
Mary University of London, London, E1 4NS, United Kingdom}
\affiliation{The Alan Turing Institute, 96 Euston Rd, London NW1 2DB, United Kingdom}
\author{Hanlin Sun}
\affiliation{School of Mathematical Sciences, Queen Mary University of London, London, E1 4NS, United Kingdom}
\author{Giacomo Rapisardi}
\affiliation{Barcelona Supercomputing Center (BSC)}
\affiliation{Departament d'Enginyeria Inform\`{a}tica i Matem\`{a}tiques, Universitat Rovira i Virgili, 43007 Tarragona -- Spain}
\author{Alex Arenas}
\affiliation{Departament d'Enginyeria Inform\`{a}tica i Matem\`{a}tiques, Universitat Rovira i Virgili, 43007 Tarragona -- Spain}

\begin{abstract}
{With the hit of new pandemic threats, scientific frameworks are needed to understand the unfolding of the epidemic.  The use of mobile apps that are able to trace contacts is of utmost importance in order to control new infected cases and contain further propagation. 
Here we present a theoretical approach using both percolation and message--passing techniques, to the role of contact tracing, in mitigating an epidemic wave. We show how the increase of the app adoption level raises the value of the epidemic threshold, which is eventually maximized when high-degree nodes are preferentially targeted. Analytical results are compared with extensive Monte Carlo simulations showing good agreement for both homogeneous and heterogeneous networks. These results are important to quantify the level of adoption needed for contact-tracing apps to be effective in mitigating an epidemic.}
\end{abstract}


\maketitle
Percolation theory \cite{barabasi2016network,newman2018networks,dorogovtsev2008critical,cohen2010complex,barrat2008dynamical} constitutes a subject of major relevance in the field of complex networks. It provides a simple mathematical 
framework which naturally applies to both networks' structural properties, (such as resilience under random damage) \cite{Albert_attack_tolerance_2000,Cohen_resilience_prl_2000,Dorogovtsev_structure_prl_2000}, and critical diffusion, (such as epidemic spreading in heterogeneous structures) \cite{Newman_spread_pre_2002,Satorras_epidemic_revmod_2015}.  As a matter of fact, even though there exists several  epidemiological models with different flavors of complexity, the arguably most popular one, \ie \  the SIR model, was found \cite{Newman_spread_pre_2002,Satorras_epidemic_revmod_2015}  to be 
mappable to a static link-percolation problem, which allowed to find analytical expressions for the epidemic threshold depending
on the underlying network topology. These results, even if they might be only an approximation of observed features in 
real epidemics, still constitute a fundamental theoretical cornerstone in the field of epidemic processes. Recently there has been an increasing interest in studying the effectiveness of track and tracing policies as a measure to contain epidemic spreading 
\cite{Ferrettieabb6936,Chinazzi395,Fraser6146,kojaku2020effectiveness}:
for instance, in \cite{kojaku2020effectiveness} the authors show how an effective contact tracing strategy in scale-free networks can reduce the probability of superspreading events, while in \cite{Ferrettieabb6936} it is claimed that a widely used contact-tracing app, combined with additional measures such as social distancing might be sufficient to stop an epidemic diffusion.\\

There are several mathematical arguments proposed in the contemporary literature to justify the above-mentioned effects, for instance in \cite{kojaku2020effectiveness} a simple generating function argument is proposed in order to compute the probability that contact tracing stops the epidemic propagation, however a solid percolation approach able to capture analytically the impact of a diffused tracing app on the non-linear aspect of epidemic spreading has not been proposed so far. In this work, we take a step forward in filling this gap by proposing a stylized model for epidemic spreading with contact-tracing and testing policies based on link percolation.

In particular, {we first consider each individual $i$, of a given contact network, to be assigned a binary variable $T_i$ representing whether or not the individual has the tracing app}. Then, we propose a modified version of the popular message-passing (MP) equations \cite{Newman_messagepass_pre_2010,karrer2014percolation,
	bianconi2018multilayer,radicchi2017redundant,Cantwell23398,PhysRevX.4.021024,LU20161,PhysRevE.94.012305}   which takes into account the following rationale. Every infected individual with probability $p$, called the {\em transmissibility} of the epidemic, transmits the disease to a  susceptible neighbor.
{An individual who has got the app, will know almost instantaneously (this is an hypothesis far from reality, but simplifies the analysis) if has been in contact with an infected individual also having the app, an she/he immediately self-isolates stopping propagation. However, if infected from an individual still not having the app, she/he will not know until symptoms appear. This can be formulated as follows:}
 individuals with the app ($T_i = 1$) can infect only if previously infected by individuals without the app ($T_i = 0$), while individuals without the app can infect regardless the $T_i$ value of their infector. By doing so we are able to derive a modified non-backtracking matrix \cite{karrer2014percolation,Krzakala20935,Martin_localization_pre_2014,moore2020predicting,rogers2015assessing} whose  largest eigenvalue determines the epidemic threshold $p_c$. Furthermore, for the case of uncorrelated networks, we are also able to derive an analytical expression for $p_c$ as a function of the average distribution of the tracing app, namely $T(k)$. Our results show that in general the more the app is diffused among the population the
higher is the value of $p_c$, meaning that the endemic state is {less likely to be achieved. Moreover we show that given
a fixed app coverage on a random network ensemble, the optimal $T(k)$ which maximizes $p_c$ corresponds to a hub-targeting strategy. By applying the message-passing algorithm to  real networks we also show that this strategy gives excellent results compared with other state-of-the-art ranking algorithm for the centrality of nodes in epidemic spreading.}

\begin{figure}[t]
  \includegraphics[width=1.0\columnwidth]{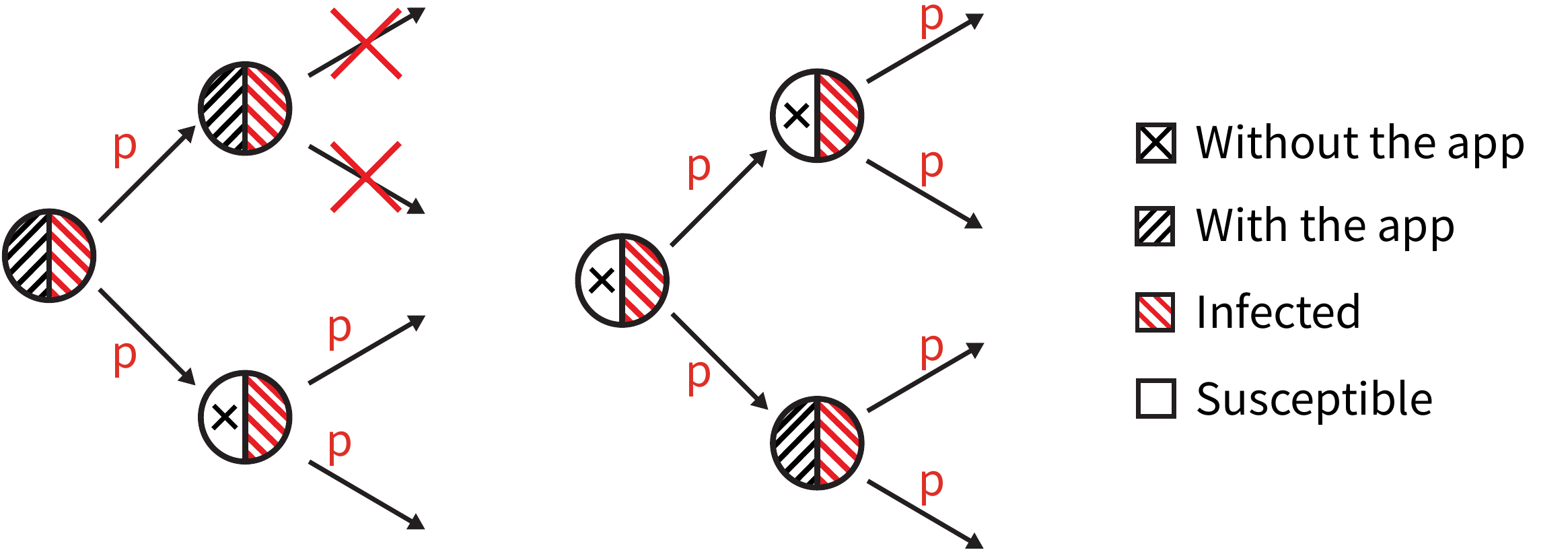}
  \caption{Sketch of the infection pathways that leads to the epidemic spreading in a population in which there are individuals that have adopted the app and individual that have not adopted the app.}
  \label{fig:model}
\end{figure}

{\it Basic model of spreading with app- } {Let us assume a contact network $G(V,E)$ formed} by $|V|=N$ individuals $i=1,2,\ldots N$, each individual $i\in V$ is assigned a variable $T_i$ indicating whether the individual has got the app $T_i=1$ or not $T_i=0$.
{Assuming the contact tracing app has immediate effect on quarantining suspicious cases, a person with the app can infect only if it is infected by a person without the app, while a person without the app can infect regardless if he has got the infection from a person with the app or without the app (see Figure $\ref{fig:model}$). Now, we propose a stochastic infection model as follows: for every link $(i,j)$ we draw a random variable $x_{ij}\in \{0,1\}$ indicating whether the eventual  contact between one infected and one susceptible node, found at the two ends of the link, leads to the infection. We parametrize this dynamic by taking $\avg{x_{ij}}=p$, where $p$ indicates the transmissibility of the epidemic.}

We can simulate the stationary state of this  spreading process on networks of arbitrary topology, i.e. including spatial networks with high clustering coefficient, by implementing the following Monte Carlo algorithm  which takes advantage of  the mapping between epidemic spreading and percolation.
We name $T-T$ the links connecting two individuals adopting the app.
These links do not contribute to the propagation of the infection to nodes other than the two connected nodes.  In order words the causal chains of infection stop when they involve a $T-T$ link.
Therefore we first consider the giant component of the link percolation process in which all the $T-T$ links are removed and all the other links are retained only if $x_{ij}=1$.  
To calculate the total fraction of infected individuals  in addition to the nodes in this giant component we  include  also the nodes with the app infected by nodes with the app. (see SM \cite{SM} for details).

{\it Message-Passing approach-}
To analytically predict the propagation of the epidemic on a network we use the powerful MP (Message-Passing) approach  \cite{lokhov2014inferring,altarelli2014bayesian,bianconi2018multilayer,karrer2014percolation,radicchi2017redundant}.
Although this approach is proven to give exact results only on locally tree-like networks, it is also well known to be very robust in the case of networks with loops, when the underlying MP algorithm converges
\cite{melnik2011unreasonable}.
In this work we adopt the MP approach and we use it to predict the phase diagram of the spreading process on network ensembles as a function of the level of adoption of the app in the population.

The considered spreading model is stochastic and has different sources of randomness that can be taken into account by different MP algorithms in which we average different level of information \cite{bianconi2018multilayer}.
The  simplest message MP can be derived   assuming to know everything about the spreading dynamics. This would entail first to know the contact   network, secondly   to know which individuals have the app, i.e. the configuration $\{T_i\}_{i\in V}$, and finally to know which links have led to an actual infection, i.e. $\{x_{ij}\}_{(i,j)\in E}$ (see SM \cite{SM} for details). 
One can  then relax the hypothesis of perfect knowledge about the epidemic process and we can consider the message passing processes in which we average over the distribution of $\{x_{ij}\}_{(i,j)\in E}$.
In this situation the outcome of the epidemic spreading is dictated by the following MP equations.
A  node $i$ spread the virus to node $j$ only with probability ${\sigma}_{i\to j}\in [0,1]$ where this message is found by the MP equation
\bea
{\sigma}_{i\to j}&=&p T_i \left[1-\prod_{\ell\in N(i)\setminus j}(1-(1-T_{\ell}){\sigma}_{\ell\to i})\right]\nonumber \\
&&+p(1-T_i) \left[1-\prod_{\ell\in N(i)\setminus j}(1-{\sigma}_{\ell\to i})\right],\label{mes2}
\eea
where $N(i)$ indicates the neighbours of node $i$. These equations directly implement the model as described in Fig. $\ref{fig:model}$. 
Moreover a node $i$ is infected with probability $\sigma_i\in [0,1]$ with
\bea
{\sigma}_{i}=\left[1-\prod_{\ell\in N^(i)}(1-{\sigma}_{\ell\to i})\right].
\eea
Therefore the expected fraction $S$ of infected individuals is given by 
\bea
S=\frac{1}{N}\sum_{i=1}^N\sigma_i.
\label{S}
\eea
This process has an epidemic threshold achieved when the maximum eigenvalue $\Lambda(\mathcal{B})$ of the modified non-backtracking matrix ${\mathcal{B}}$  is equal to one, i.e.
\bea
\Lambda(\mathcal{B})=1.
\label{Lambda}
\eea 
The modified non-backtracking matrix ${\mathcal{B}}$ for this algorithm is defined in terms of the non-backtracking matrix ${\mathcal{A}}$ of the network  as 
\bea
{\mathcal{B}}_{ \ell i \to ij}=p (1-T_i T_{\ell})\mathcal{A}_{\ell i \to ij}.
\label{nonback}
\eea
Here ${ \mathcal{A}}$ \cite{karrer2014percolation} has elements
\bea
\mathcal{A}_{  \ell i  \to ij}=a_{\ell i}a_{ij}(1-\delta_{\ell j}),
\eea
where ${\bf a}$ is the adjacency matrix of the network and $\delta_{rs}$ is the Kronecker delta.
 Equations (\ref{Lambda}) and (\ref{nonback}) clearly show that the epidemic threshold is dictated essentially by the non-backtracking matrix of the network where we have removed all the $T-T$ links.

We can also average over the probability distribution of $\{T_i\}_{i\in V}$. 
Specifically we can assume  that $\overline{T_i}$ (the $\overline{\ldots}$ indicates the average over the probability distribution of $\{T_i\}_{i\in V}$) is only a function of the node degree, i.e. $\overline{T_i}=T(k_i)$.   This is a minimal assumption that allows to derive analytical calculations, however we note that the adoption of the app might depend on an additional social contagion process of awareness behavior in a  scenario close to the one proposed in  Ref.~\cite{granell2014competing}.  In order to mimic these alternative scenarios in the SM \cite{SM} we have considered the cases in which the adoption of the app depends on either the eigenvector centrality or the non-backtracking centrality of the nodes. 

For formulating the MP  algorithm in the case in which we assume to known only the function $T(k)$, the transmissibility $p$, and the actual contact network, we consider for every ordered pair of linked nodes $(i,j)$ the two messages indicating the probability that node $i$ infects node $j$ given that node $i$ has adopted ($\hat{\sigma}^T_{i\to j}$) or not adopted ($\hat{\sigma}^N_{i\to j}$) the app.
These two messages are given by  
\bea
\hat{\sigma}^T_{i\to j}&=&\overline{T_i\sigma_{i\to j}},\nonumber \\
\hat{\sigma}^N_{i\to j}&=&\overline{(1-T_i)\sigma_{i\to j}}.
\eea
The MP equations for these  messages can be obtained by averaging the MP Eqs.(\ref{mes2}) over all the configuration $\{T_i\}_{i\in V}$ and read: 
\bea
\hat{\sigma}^N_{i\to j}&=&p(1-T(k_i))\left[1-\prod_{\ell\in N(i)\setminus j}(1-\hat{\sigma}^N_{\ell\to i}-\hat{\sigma}^T_{\ell\to i})\right]\nonumber\\
\hat{\sigma}^T_{i\to j}&=&pT(k_i)\left[1-\prod_{\ell\in N(i)\setminus j}(1-\hat{\sigma}^N_{\ell\to i})\right].
\label{mes_av1}
\eea
The probability that node $i$ is infected ${{\sigma}}_i$ is given by 
\bea
{{\sigma}}_{i}=\left[1-\prod_{\ell\in N^(i)}(1-\hat{\sigma}^N_{\ell\to i}-\hat{\sigma}^T_{\ell\to i})\right],
\label{mes_av2}
\eea
while the expected fraction $S$ of infected nodes is given by Eq. (\ref{S}).
In this case the relevant matrix $\mathcal{B}$ determining the epidemic threshold given by  Eq. (\ref{Lambda}) is (see SM \cite{SM} for details)
\bea
{\mathcal{B}}_{ \ell^{\prime} \ell\to ij}&=&p[1-T({k_i})]\delta_{\ell i}\mathcal{A}_{ \ell^{\prime} i \to ij}\nonumber \\
&+&p^2[1-T({k_i})]T({k_\ell})\mathcal{A}_{ \ell^{\prime}\ell  \to \ell i}\mathcal{A}_{ \ell i \to ij}.
\eea
 
Finally we consider the case in which we do not have perfect knowledge about the network itself and can perform the average over an uncorrelated network ensemble. In this case we have two equations: one for $S^{\prime}_N$
and one for $S^{\prime}_T$, indicating the probability that by following a link  we reach  an infected  individual without the  app or with the app respectively. These equations (see SM \cite{SM} for details of the derivation) read,
\bea
S^{\prime}_N&=&p\sum_{k}\frac{kP(k)}{\avg{k}}(1-T(k))\left[1-(1-S^{\prime}_N-S^{\prime}_{T})^{k-1}\right],\nonumber \\
S^{\prime}_T&=&p\sum_{k}\frac{kP(k)}{\avg{k}}(T(k))\left[1-(1-S^{\prime}_N)^{k-1}\right].
\eea
Here $T(k)$ indicates the probability that a node of degree $k$ gets the app.
The probability that a random node gets the infection is given by
\bea
S&=&\sum_{k}{P(k)}\left[1-(1-S^{\prime}_T-S^{\prime}_N)^{k}\right],
\eea
The transition is achieved for 
\bea
p_c={\min}\left(1,\frac{1}{2\kappa_T}\left[-1+\sqrt{1+4\frac{\kappa_T}{\kappa_N}}\right]\right).
\label{pc}
\eea
where 
\bea
\kappa_N&=&\frac{\avg{k(k-1)(1-T(k))}}{\avg{k}}.
\nonumber \\
\kappa_T&=&\frac{\avg{k(k-1)T(k)}}{\avg{k}}.
\eea
{\it Optimization -}

\begin{figure*}
	\includegraphics[width=17.2cm]{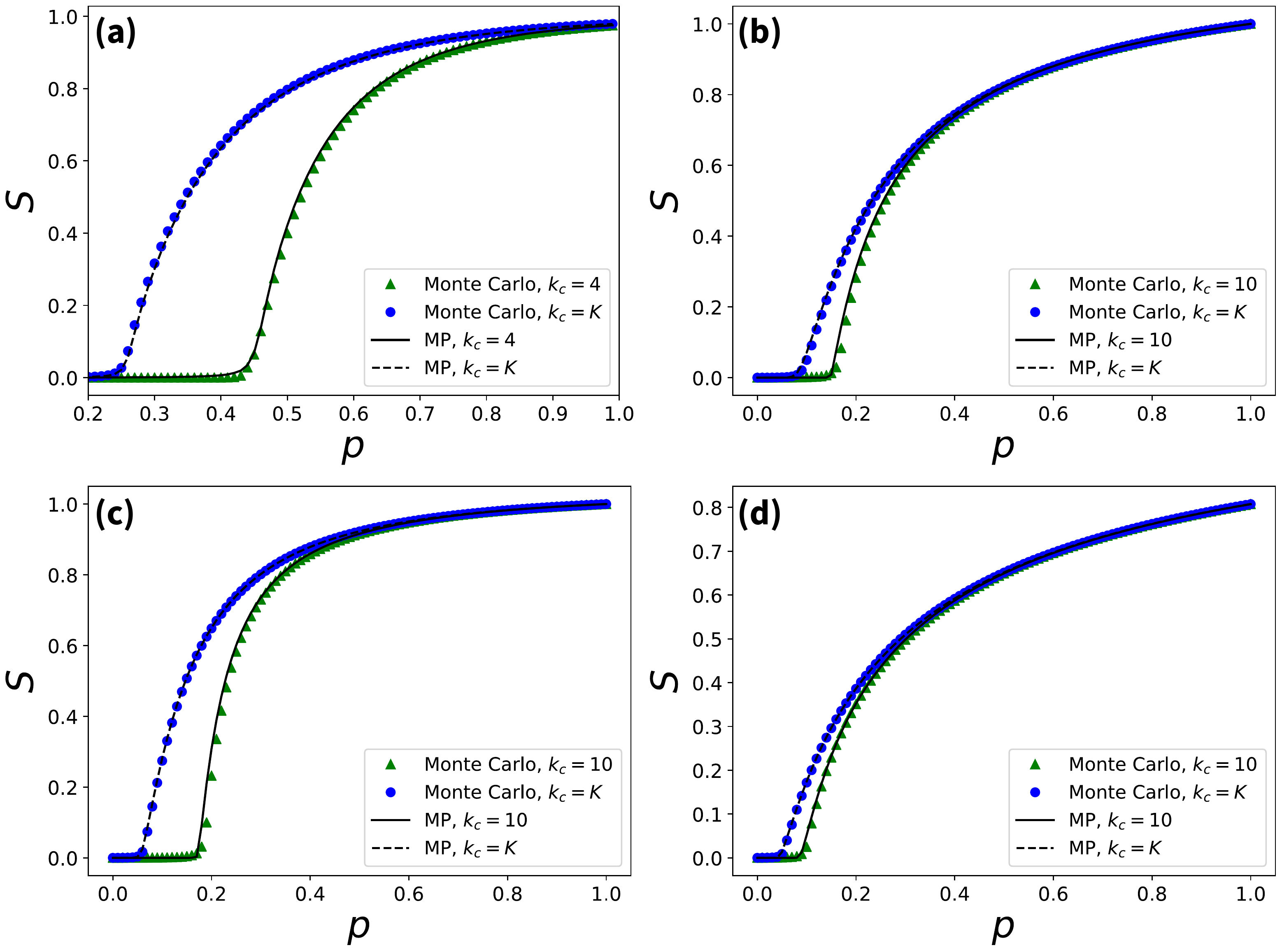}
	\caption{(Color online) The fraction of infected nodes $S$ is plotted versus $p$ for several networks. The results obtained by averaging the Monte Carlo simulations of the  configurations $\{T_i\}_{i\in V}$ and $\{x_{ij}\}_{(i,j)\in E}$  are compared with the results of the MP algorithm defined by Eqs. (\ref{mes_av1}) and Eq. (\ref{mes_av2}), where $T(k)$ is given by Eq. \eqref{opt} with $\alpha = 0$ and $k_c$ as indicated in the legend of each panel. The value $K$ in all panels corresponds to the largest degree of the network and therefore corresponds to the case of no app coverage. \textbf{(a)} Poisson network with $N=5\times 10^4$ nodes and average degree $\lambda = 4$. \textbf{(b),(c),(d)} Friendship networks from the music streaming site Deezer in the countries of Romania ($N = 41773$), Hungary ($N = 47538$) and Croatia ($N = 54573$) respectively \cite{Rozemberczki_2019}.
	}
	\label{fig:message_passing}
\end{figure*}

  The formula for $p_c$, provided by  Eq.~(\ref{pc}),  is an increasing function of $\kappa_T$ so in order to {maximize} $p_c$ we need to maximize $\kappa_T$.
Under the $\mathbb{L}_1$ norm 
\bea
\sum_{k}P(k)T(k)=\mathcal{T}.
\label{L1}
\eea
This optimization problem gives the   discrete Heaviside step function
\begin{equation}\label{opt}
\tilde{T}(k)=\theta(k-k_c,\alpha)
\end{equation}
taking the value $0\leq \alpha=\mathcal{T}-\sum_{k>k_c}P(k)<1$ at $k=k_c$.
Therefore the optimal solution is to have all nodes of degree $k>k_c$ with $100\%$ app adoption and the node with exactly $k=k_c$ with the maximal adoption allowed by the constraint in Eq. (\ref{L1}).
For this choice of $T(k)$ we have checked the validity of the proposed message passing theory by comparing the results obtained by a direct implementation of the {Monte Carlo} algorithm predicting the fraction of nodes affected by the epidemics with the results of the  MP algorithm defined in Eq. (\ref{mes_av1}), (\ref{mes_av2})  finding an excellent agreement between the two, for both real and synthetic networks (see Figure \ref{fig:message_passing}).

\begin{figure}[t]
  \includegraphics[width=8.6cm]{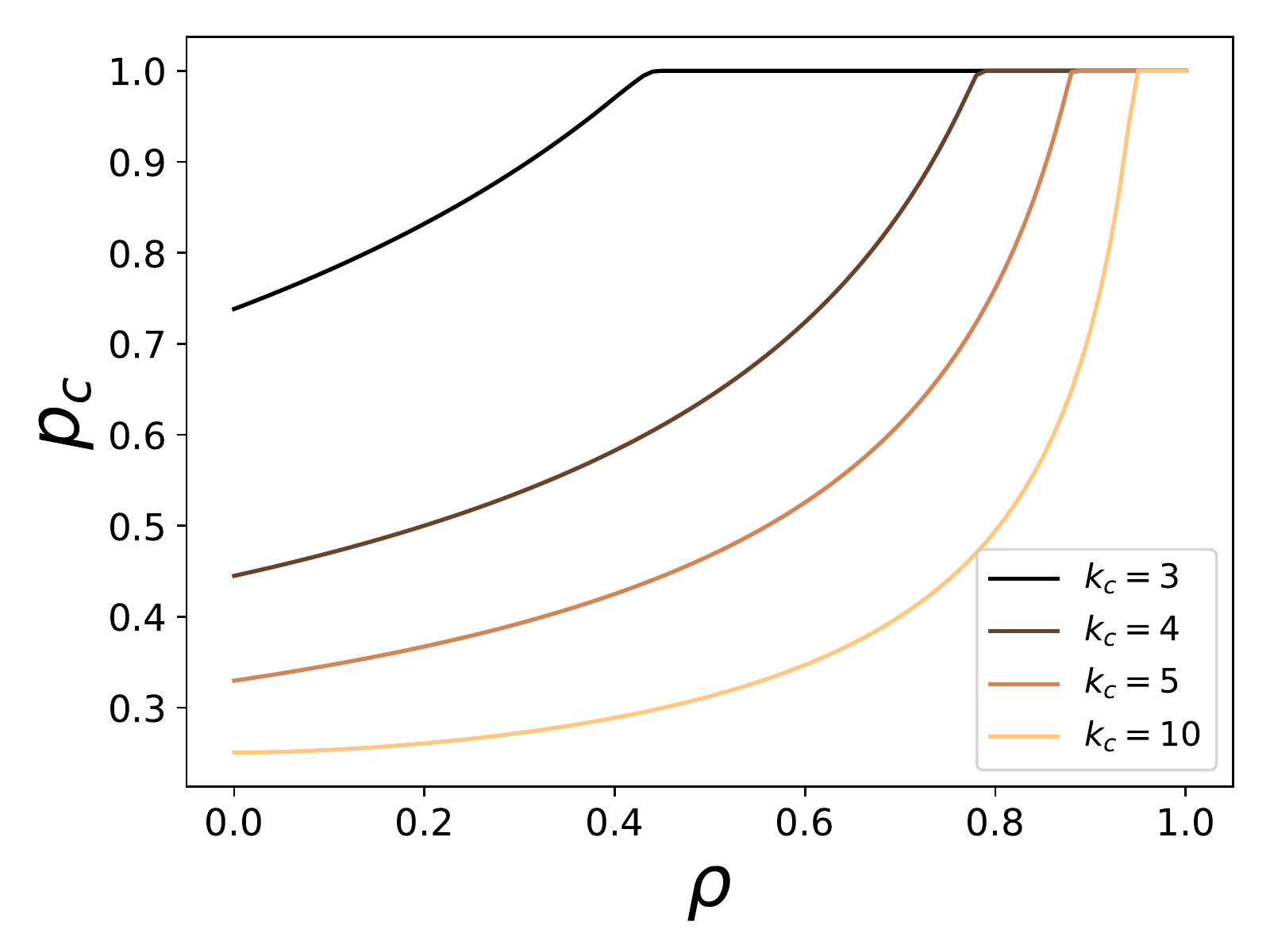}
  \caption{ (Color online) The phase diagram of the epidemic model mitigated by the adoption of the app is shown for a Poisson network of $N = 10^4$ nodes with average degree $\lambda=4$. Here  $T(k)$ is given by Eq. (\ref{t_k_rho}) with $\alpha=0$. The epidemic threshold $p_c$ is plotted as a function of $\rho$ for different values of the cutoff $k_c$.}
  \label{fig:phase_diagram}
\end{figure}

{\it Improvement on $p_c$-}
Equation \eqref{opt} tells us that in an uncorrelated random network, given a fixed app coverage $\mathcal{T}$, the best strategy in order to maximally delay the percolation transition is given by targeting the hubs. In order to verify the optimality of Eq. \eqref{opt} when compared to different strategies, we considered the more general form of $T(k)$ given by:
\bea
	T(k) = \rho + (1-\rho)\theta(k-k_c,\alpha),
\label{t_k_rho}
\eea
where $\theta(k-k_c)$ is the discrete Heaviside step function taking the value $\alpha$ at $k=k_c$, and $\rho \in [0,1]$ denotes a uniform fraction of individuals adopting
the app. Thanks to Eq. \eqref{t_k_rho} we are able to interpolate between a purely random strategy obtained by taking the limit
$k_c \rightarrow \infty$ and the optimal strategy given in the limit $\rho \rightarrow 0$. It is straightforward to check that under
the constraint defined in Eq. \eqref{L1} we have respectively $\lim_{k_c \rightarrow \infty} T(k) = \mathcal{T}$ and 
$\lim_{\rho \rightarrow 0} T(k) = \tilde{T}(k)$. \\
We have used Eq. (\ref{pc}) to investigate the phase diagram (characterized by the epidemic threshold $p_c$) of a Poisson network as a function of $\rho$ and $k_c$ (see Figure $\ref{fig:phase_diagram}$). We observe that a diffused adoption of the app can significantly increase $p_c$, which happens when $\rho$ increases or when $k_c$ decreases.
 
\begin{figure}[t]
	\includegraphics[width=8.6cm]{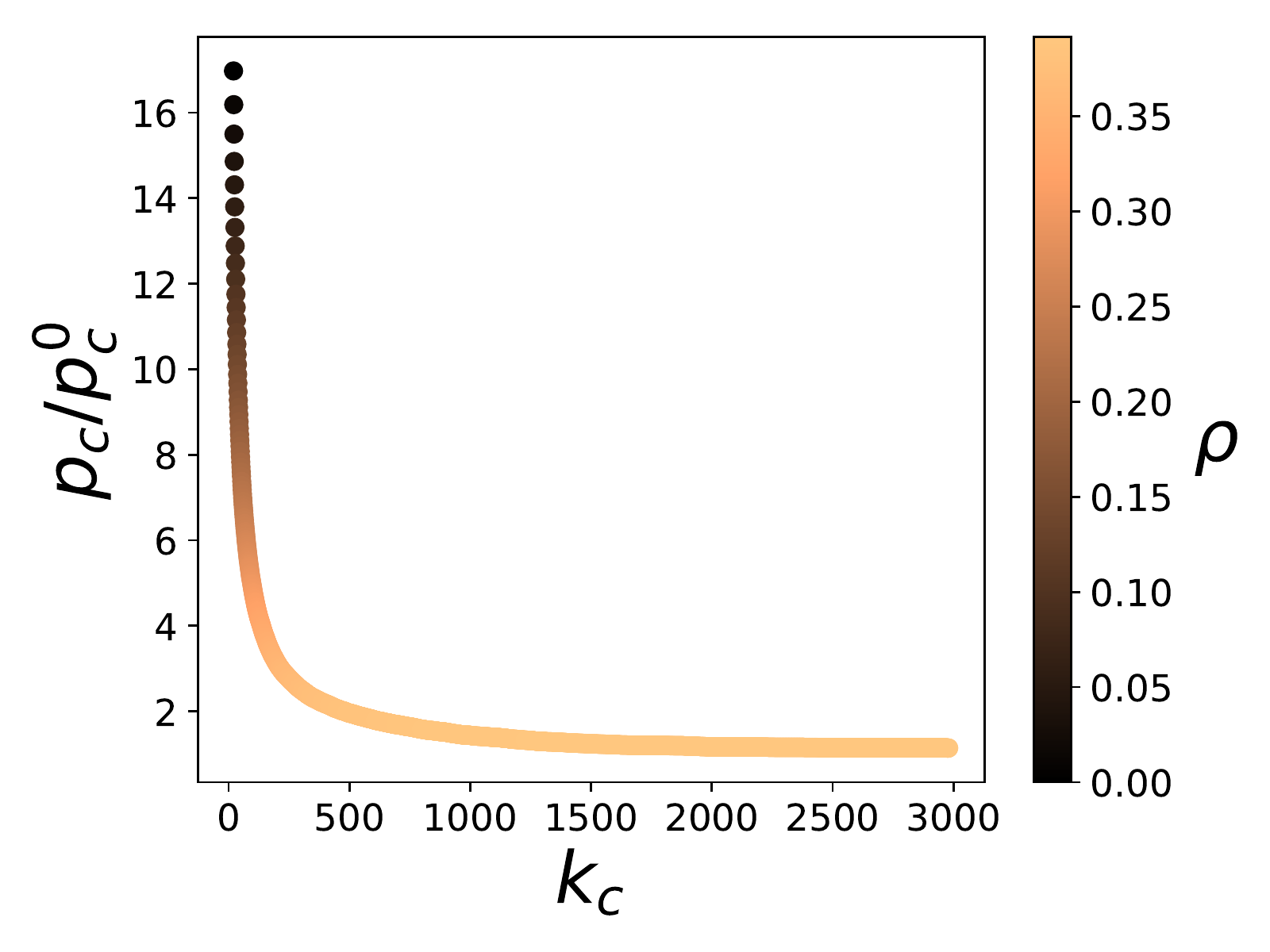}
	\caption{ (Color online) Relative increase of $p_c$ computed from Eq. \eqref{pc} on the Livemocha social-network ($N\sim 104\times 10^3$ nodes, $E \sim 2\times 10^6$ edges) \cite{konect:socialcomputing}, where $T(k)$ is given by Eq. \eqref{t_k_rho} under the constraint \eqref{L1}, and $p_c^0 = \avg{k}/\avg{k(k-1)}$ represents the value of the percolation threshold in the absence of app coverage (which can be obtained from Eq. \eqref{pc} in the limit $\kappa_T \rightarrow 0$). Here $p_c^{0} = 0.00306$, while the app coverage is fixed at $\mathcal{T} = 0.39175$, corresponding to an optimal $\tilde{T}(k)$ with $k_c = 20$ and $\alpha = 1$. The plot shows that for this particular value of $\mathcal{T}$, corresponding to
		$\sim$40\% of the nodes having the app, the optimal
		distribution is reached at $\rho = 0$ and corresponds to a $\sim$17-fold increase of $p_c$, whereas in the case of a purely random strategy, obtained at $\rho = \mathcal{T}$, the increase of $p_c$ is $\sim$1.2-fold.
	}\label{fig:pc_livemocha}

\end{figure}

To show, in a particular example, the increase of $p_c$ due to the adoption of the app, we consider the real dataset Livemocha social-network \cite{konect:socialcomputing}.  As we can see from Fig. \ref{fig:pc_livemocha} the random adoption strategy, achieved when $k_c = k_{max}$, yields a very small increase in the value of $p_c$ compared to the optimal distribution, corresponding to $\rho = 0$. Therefore in a scenario of limited resources, represented by the constraint defined in Eq. \eqref{L1}, the optimal strategy corresponds to distribute the app from higher-degree nodes to lower-degree ones until the resources are exhausted. The resulting increase in $p_c$ computed according to Eq. \eqref{pc} is quite dramatic and non trivial, for instance from Fig. \ref{fig:pc_livemocha} we read that if the app is optimally
	distributed among $\sim$40\% of the population the increase of $p_c$ is $\sim$17-fold, while if the same percentage is covered randomly the increase is $\sim$1.2-fold.
This optimization principle is obtained under the assumption that the adoption of the app is dictated by the degree of the nodes. However in a real scenario this hypothesis might appear too restrictive. Devising an ad-hoc optimization algorithm similar to the ones proposed in \cite{morone2015influence,lokhov2014inferring,altarelli2014bayesian} is beyond the scope of this Letter. However, in order to check how the obtained optimal strategy compare with other possible mechanisms driving the adoption of the app in the SM we show that targeting the hubs remains a very good strategy also if compared to targeting the high eigenvector centrality nodes or the high non-backtracking centrality nodes \cite{moore2020predicting,rogers2015assessing} in a number of real datasets.   

{\it Conclusions-}
{In this work we provide a message-passing theory able to predict the epidemic threshold of disease spreading among a population which has the option of adopting a tracing app. The simplicity of our model allows us to derive a simple analytical estimate for the epidemic threshold and leaves plenty of room for taking into account more complex and realistic factors. For instance, we assumed that the tracing app is perfect, however the we can relax this assumption in order to allow also for imperfect tracing and isolation. Another interesting follow up for the model could be the introduction of a time dimension similar to the one proposed in \cite{moore2020predicting}, in order to assess how the modified non-backtracking matrix presented in Eq.\eqref{nonback} affects not only the percolation threshold itself, but also the speed of the epidemic.\\
The proposed stylized mathematical framework can overall be useful to assess the expected impact of contact-tracing apps in the course of an epidemics if adopted correctly. The compartmental epidemic model used is the classical SIR, and do not pretend to be a model fitted for the current pandemic of COVID-19, however the physical intuition we grasp from the presented analysis may prove fundamental to prescribe the best targeting strategy for app adoption, as well as it captures the highly non-linear effect on the reduction of the incidence provided by a certain fraction of adoption.
	Our preliminary results show both numerically and theoretically that the adoption of the app by a large fraction of the population increases the value of the epidemic threshold. In case of uncorrelated networks we are able to derive a closed analytic expression for $p_c$ which 
	depends on both the network degree-distribution $P(k)$ and the average app distribution $T(k)$. Thanks to this expression we finally
	prove  in a constrained-resources scenario that the value of $p_c$ is maximized when high-degree nodes are preferentially targeted.
	Our results show that an optimal targeting gives rise to a dramatic increase in the value of $p_c$ when compared to a strategy
	in which the same amount of resources is uniformly distributed. The more randomly the app is diffused among the population the less is the increase in the percolation threshold, or equivalently, the less the app has the power of mitigating the epidemics.
	Overall our results show that even if the adoption of a tracing app has the effect of preventing an epidemic wave, the same level of adoption
	can be optimally distributed by taking into account the heterogeneity of the population contact network in order to obtain a mitigation effect which is significantly higher.
	} \\

	\begin{acknowledgments}
AA acknowledges support by Ministerio de Econom\'{\i}a y Competitividad (grant FIS2015-71582-C2-1), Generalitat de Catalunya (grant 2017SGR-896), and Universitat Rovira i Virgili (grant 2017PFR-URV-B2-41), ICREA Academia and the James S.\ McDonnell Foundation (grant \#220020325). HS acknowledges funding by the Chinese Scholarship Council.
\end{acknowledgments}

\bibliographystyle{apsrev4-1}

\bibliography{biblio_track}

\clearpage
\renewcommand\theequation{{S-\arabic{equation}}}
\renewcommand\thetable{{S-\Roman{table}}}
\renewcommand\thefigure{{S-\arabic{figure}}}
\setcounter{equation}{0}
\setcounter{figure}{0}
\setcounter{section}{0}

\onecolumngrid
\appendix

\begin{center}
{\Large{\bf SUPPLEMENTARY MATERIAL}}
\end{center}
\section*{Mapping of epidemic spreading to percolation problem}
We assume that the  network $G=(V,E) $ of contacts is formed by $N=|V|$ individuals $i=1,2,\ldots N$. Each individual is assigned a variable $T_i$ indicating whether the individual has adopted  the app ($T_i=1$) or not ($T_i=0$).
Assuming that the effect of track and tracing is immediate, a node with the app can infect only if its is infected by a node without the app, while a node without the app can infect regardless the state of the infector node.
For every link $(i,j)\in E$ we draw a random binary  variable $x_{ij}\in \{0,1\}$ indicating  weather the eventual contact between one infected an one susceptible node find at the two ends of the link leads to the infection ($x_{ij}=1$), or not ($x_{ij}=0$). Here we assume that  the average of $\avg{x_{ij}}$  is given by the trasmissibility $p$, i.e. $\avg{x_{ij}}=p$.

In order to find which are the nodes infected in the epidemic outbreak we adopt the following algorithm that uses the mapping of the stationary state of epidemic to percolation \cite{Newman_spread_pre_2002}.

\begin{itemize}
\item{\it Pre-processing of the connections-}
We call $T-T$ the links connecting two individuals both adopting  the app.
These links do not contribute to the propagation of the infection to nodes other than the two connected nodes. Therefore we initially remove from the network all $T-T$ links.
Specifically we associate to each link $(i,j)$  the variable $y_{ij}\in\{0,1\}$ defined as
\bea
y_{ij}=x_{ij}(1-T_iT_j),
\eea
indicating whether the link contributes or not the spread of the disease in the network (excluding the two nodes $(i,j)$ of the link).
\item{\it Percolation process-}
We find the nodes in the giant component of the resulting percolation problem. We assign to each node the indicator variable
$m_i\in \{0,1\}$ indicating if node $i$ belongs or not to the giant component of the network with  links according to the indicator function $y_{ij}$.
The nodes with $m_i$ are nodes that are infected by chain of contacts in which there we can never find two consecutive infected nodes with the app. 
\item{\it Calculation of the fraction of infected individuals-}
In order to calculate the total fraction of infected individual we need to include in addition to the nodes with $m_i=1$ also the nodes with the app infected by nodes with the app. Therefore we define an indicator function $\sigma_i$ which will indicate for  each individual if it is  infected ($\sigma_i=1$) or not  ($\sigma_{i}=0$). The value of $\sigma_i$ can be evaluated  according to the boolean rule 
\bea
\sigma_i=m_i+(1-m_i)(1-\prod_{j\in N(i)}(1-m_jT_jT_ix_{ij})),
\eea
\end{itemize}
\section*{Message passing algorithms for epidemic spreading in a population partially adopting the app}
In this section we discuss the message passing algorithms \cite{karrer2014percolation,bianconi2018multilayer} that can be used to predict the outcome of the epidemic spreading studied in this work. We will first assume to have full knowledge about the configuration $\{T_i\}_{i \in V}$ and $\{x_{ij}\}_{(i,j)\in E}$ and subsequently we will relax this strong assumption by assuming to know only the value of the transmissibility $p$ fixing the expectation $\avg{x_{ij}}=p$. Finally we will relax further our assumptions and we will consider the case in which the configuration  $\{T_i\}_{i \in V}$ is also not known  exactly and only the expectations $\overline{T_i}=T(k_i)$ where $k_i$ is the degree of the generic node $i$ is known. 

In the first case in which to known exactly the configuration $\{T_i\}$ and $\{x_{ij}\}$  the message passing algorithm on a locally tree-like network predict that  a  node $i$ spread the virus to node $j$ only if  $\tilde{\sigma}_{i\to j}=1$. If   node $i$ has the app,i.e.  $T_i=1$, the message $\tilde{\sigma}_{i\to j}$ is one i.e. $\tilde{\sigma}_{i\to j}=1$, if node $i$ has been infected by at least a neighbour node  without the app and $x_{ij}=1$, otherwise $\tilde{\sigma}_{i\to j}=0$. If node $i$ does not have the app, i.e.  $T_i=0$ the message $\tilde{\sigma}_{i\to j}$ is one, i.e.  $\tilde{\sigma}_{i\to j}=1$ if node $i$ has been infected by at least a neighbour node  and $x_{ij}=1$,  otherwise $\tilde{\sigma}_{i\to j}=0$.  
Therefore the message passing algorithm reads
\bea
\tilde{\sigma}_{i\to j}&=&x_{ij} T_i \left[1-\prod_{\ell\in N(i)\setminus j}(1-(1-T_{\ell})\tilde{\sigma}_{\ell\to i})\right]+x_{ij}(1-T_i) \left[1-\prod_{\ell\in N(i)\setminus j}(1-\tilde{\sigma}_{\ell\to i})\right],\nonumber 
\eea
where $N(i)$ indicates the neighbours of node $i$.
Moreover the  function $\sigma_i$ indicating if a node $i$ is infected $\tilde{\sigma}_i=1$ or not $\tilde{\sigma}_i=0$ is given by 
\bea
\tilde{\sigma}_{i}=\left[1-\prod_{\ell\in N^(i)}(1-\tilde{\sigma}_{\ell\to i})\right].
\label{eq:mes1}
\eea
If follows that the epidemic threshold is determined   by the equation 
\bea
\Lambda(\mathcal{B})=1.
\label{LambdaS}
\eea 
Here $\Lambda(\mathcal{B})$  is the maximum eigenvalue of the corrected non-backtracking matrix $\mathcal{B}$ of elements
\bea
{\mathcal{B}}_{ \ell i\to ij}=x_{ij} (1-T_i T_{\ell})\mathcal{A}_{ \ell i\to ij},
\eea
with $\mathcal{A}$ defined in terms of the adjacency matrix of the network ${\bf a}$ as 
\bea
\mathcal{A}_{  \ell i  \to ij}=a_{\ell i}a_{ij}(1-\delta_{\ell j}).
\eea
This algorithm should be modified if we do not have access to the full configuration of $\{x_{ij}\}_{(i,j)\in E}$.
In this case we assume to  know only the trasmissibility of the disease $p=\avg{x_{ij}}$. In this case the messages are real values  ${\sigma}_{i\to j}\in [0,1]$ and indicate the probability that node $i$ infects node $j$.
By averaging the message passing equations over all possible configuration $\{x_{ij}\}$ (see \cite{bianconi2018multilayer} for a overview of this technique) at fixed value of the transmissibility of the infection $p$ we obtain the message passing algorithm 
\bea
{\sigma}_{i\to j}&=&p T_i \left[1-\prod_{\ell\in N(i)\setminus j}(1-(1-T_{\ell}){\sigma}_{\ell\to i})\right]+p(1-T_i) \left[1-\prod_{\ell\in N(i)\setminus j}(1-{\sigma}_{\ell\to i})\right],\label{mes2b}
\eea
where $N(i)$ indicates the neighbours of node $i$.
Moreover a node $i$ is infected with probability $\sigma_i$ given by 
\bea
{\sigma}_{i}=\left[1-\prod_{\ell\in N^(i)}(1-{\sigma}_{\ell\to i})\right].
\eea
The epidemic threshold is always determined by Eq.(\ref{LambdaS}) with $\mathcal{B}$ taking the expression
\bea
{\mathcal{B}}_{ \ell i\to ij}=p (1-T_i T_{\ell})\mathcal{A}_{ \ell i\to ij}.
\eea
In order to model different scenarios corresponding to different adoption patterns of the app we might also assume that the configuration $\{T_i\}_{i\in V}$ is not known exactly and we have only access to the probability that a node adopt the app. Assuming that this probability is a function of the degree of the nodes we have $\overline{T_i}=T(k_i)$
with  $T(k)$ describing the probability that a node of degree $k$ adopts the app.
For formulating the message passing algorithm in this case we consider for every ordered pair of linked nodes $(i,j)$ the two messages 
\bea
\hat{\sigma}^T_{i\to j}&=&\overline{T_i\sigma_{i\to j}},\nonumber \\
\hat{\sigma}^N_{i\to j}&=&\overline{(1-T_i)\sigma_{i\to j}},
\eea
indicating the probability that node $i$ infects node $j$ given that node $i$ has adopted $\hat{\sigma}^T_{i\to j}$ or not adopted $\hat{\sigma}^N_{i\to j}$ the app. 
The message passing equations for these  messages can be obtained averaging the message passing Eqs.(\ref{mes2b}) over all the configuration $\{T_i\}_{i\in V}$ and read 
\bea
\hat{\sigma}^T_{i\to j}&=&pT(k_i)\left[1-\prod_{\ell\in N(i)\setminus j}(1-\hat{\sigma}^N_{\ell\to i})\right]\nonumber \\
\hat{\sigma}^N_{i\to j}&=&p(1-T(k_i))\left[1-\prod_{\ell\in N(i)\setminus j}(1-\hat{\sigma}^N_{\ell\to i}-\hat{\sigma}^T_{\ell\to i})\right].
\label{mes_av1b}
\eea
The probability that node $i$ is infected ${\tilde{\sigma}}_i$ is given by 
\bea
{\tilde{\sigma}}_{i}=\left[1-\prod_{\ell\in N^(i)}(1-\hat{\sigma}^N_{\ell\to i}-\hat{\sigma}^T_{\ell\to i})\right].
\label{mes_av2b}
\eea
The critical threshold is obtained by linearising the message passing Eqs. (\ref{mes_av1b}), getting 
\bea
\hat{\sigma}^T_{i\to j}&=&pT(k_i)\sum_{\ell\in N(i)}\mathcal{A}_{\ell i\to ij}\hat{\sigma}^N_{\ell\to i},\nonumber \\
\hat{\sigma}^N_{i\to j}&=&p(1-T(k_i)) \sum_{\ell\in N(i)}\mathcal{A}_{\ell i\to ij}(\hat{\sigma}^N_{\ell\to i}+\hat{\sigma}^T_{\ell\to i}).
\eea 
In this way by solving this linear system of equations we get 
\bea
\hat{\sigma}^T_{\ell\to i}&=&pT(k_{\ell})\sum_{\ell^{\prime}\in N(\ell)}\mathcal{A}_{\ell^{\prime} {\ell}\to \ell i}\hat{\sigma}^N_{\ell^{\prime}\to \ell},\nonumber \\
\hat{\sigma}^N_{i\to j}&=&p(1-T(k_i)) \sum_{\ell\in N(i)}\mathcal{A}_{\ell i\to ij}\hat{\sigma}^N_{\ell\to i} +pT(k_{\ell})\sum_{\ell^{\prime}\in N(\ell)}\mathcal{A}_{\ell^{\prime} {\ell}\to \ell i}p(1-T(k_i)) \sum_{\ell\in N(i)}\mathcal{A}_{\ell i\to ij}\hat{\sigma}^N_{\ell^{\prime}\to \ell}.
\eea
Therefore we obtain that the critical point is characterized the Eq.(\ref{LambdaS})
where ${\mathcal{B}}$ is given by 
\bea
{\mathcal{B}}_{ \ell^{\prime} \ell\to ij}=p\delta_{\ell,i}\mathcal{A}_{ \ell^{\prime} i \to ij}(1-T_{k_i})+p^2\mathcal{A}_{ \ell^{\prime}\ell  \to \ell i}T_{k_\ell}\mathcal{A}_{ \ell i \to ij}(1-T_{k_i}).
\eea

\section*{Ensemble approach}
In this section we show the derivation of the epidemic threshold $p_c$ in the case in which we do not know exactly the structure of the contact network, i.e. we only known that the network is a random uncorrelated network with a given degree distribution $P(k)$ and  we know only the statistical properties of the configurations $\{T_i\}_{i\in V}$ and $\{x_{ij}\}_{(i,j)\in E}$.
We consider the variables $S^{\prime}_T$
and $S^{\prime}_N$ indicating the probability that by following a link  we reach  an infected  individual with app or without app respectively. 
By averaging the message passing Eqs. (\ref{mes_av1b}) over the network ensemble  we get
\bea
S^{\prime}_T&=&p\sum_{k}\frac{kP(k)}{\avg{k}}(T(k))\left[1-(1-S^{\prime}_N)^{k-1}\right]\nonumber \\
S^{\prime}_N&=&p\sum_{k}\frac{kP(k)}{\avg{k}}(1-T(k))\left[1-(1-S^{\prime}_N-S^{\prime}_T)^{k-1}\right],
\label{Ensembleb}
\eea
where $T(k)$ indicates the probability that a node of degree $k$ adopt the app.
The probability that a random node gets the infection is given by
\bea
S&=&\sum_{k}{P(k)}\left[1-(1-S^{\prime}_T-S^{\prime}_N)^{k}\right].
\eea
The  system of Eqs. (\ref{Ensembleb}) can be written as 
\bea
S^{\prime}_{T}&-&p\sum_{k}\frac{kP(k)}{\avg{k}}(T(k))\left[1-(1-S^{\prime}_N)^{k-1}\right]=0,\nonumber \\
S^{\prime}_N&-&p\sum_{k}\frac{kP(k)}{\avg{k}}(1-T(k))\left[1-(1-S^{\prime}_N-S^{\prime}_T)^{k-1}\right]=0.
\eea
The Jacobian of this system of equations is given by   
\bea
{\bf J}=\left(\begin{matrix}1&-p\kappa_T\\
-p\kappa_N& 1-p\kappa_N\end{matrix}\right),
\eea
where 
\bea
\kappa_N&=&\frac{\avg{k(k-1)(1-T(k))}}{\avg{k}}.
\nonumber \\
\kappa_T&=&\frac{\avg{k(k-1)T(k)}}{\avg{k}}.
\eea
Imposing that the determinant of the Jacobian is zero we obtain that the  transition is achieved for 
\bea
p_c=\min\left(1,\frac{1}{2\kappa_T}\left[-1+\sqrt{1+4\frac{\kappa_T}{\kappa_N}}\right]\right).
\label{pcb}
\eea
\section*{Numerical validation of the theoretical predictions}

We have validated the proposed message passing framework by conducting extensive numerical results using the three message passing algorithms and the MonteCarlo simulations.
We considered the choice
  \bea
	T(k) = \rho + (1-\rho)\theta(k-k_c,\alpha),
\label{t_k_rhob}
\eea
where $\theta(k-k_c)$ is the discrete Heaviside step function taking the value $\alpha$ at $k=k_c$, and $\rho \in [0,1]$ denotes a uniform fraction of individuals adopting the app.

\begin{figure}[t]
	\includegraphics[width=1.0\columnwidth]{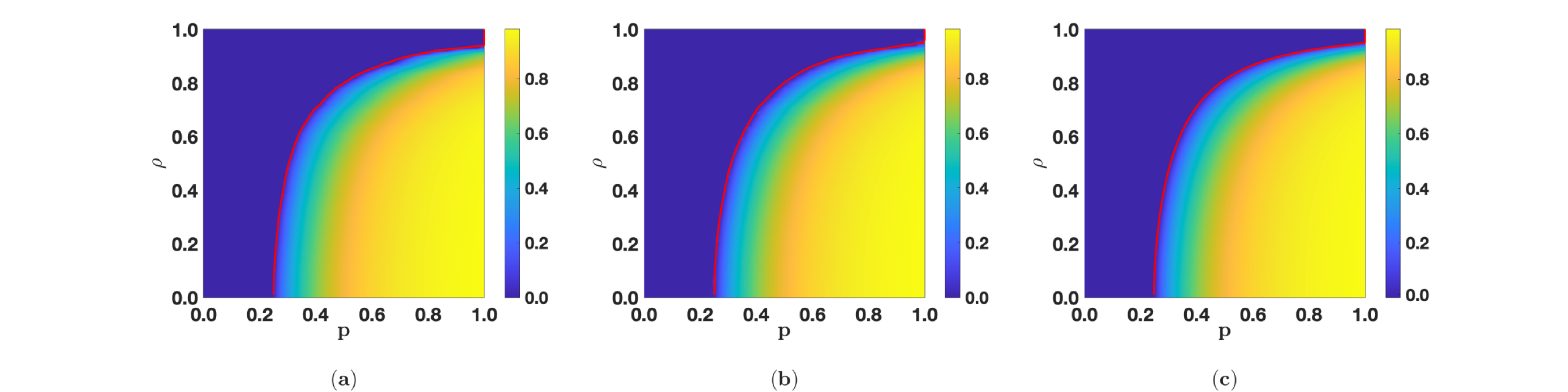}
	\caption{(Color online)The phase diagram of the epidemic model is shown by plotting the fraction $S$ of infected individuals $S$ obtained using the three different message passings in the plane $(p,\rho)$ for a $N=10^4$-node Poisson network with $\lambda=4$. Panel (a) shows the results obtained with the message passing algorithm using the exact known configuration $\{T_i\}$ and  $\{x_{ij}\}$ (Eq.\eqref{mes_av1b}),  panel (b) shows the results obtained with the message passing algorithm using exact known configuration $\{x_{ij}\}$ and transmissibility $p=\langle x_{ij}\rangle$ (Eq. \eqref{mes2b}); finally panel  (c) shows the results obtained with the message passing algorithm using transmissibility $p=\langle x_{ij}\rangle$ and probability of adopting the app $T(k)$ (Eq. \eqref{eq:mes1}). The solid (red) lines indicate the epidemic threshold predicted by Eq. (\ref{pcb}). }\label{fig:1S}
\end{figure}

The phase diagrams obtained using   the three different message passing algorithms are consistent. In particular when these algorithms are   applied to a  network drawn from a network ensemble they give results whose differences vanishes in the large network limit. To show evidence of this result, in Figure $\ref{fig:1S}$ we compared the phase diagram obtained using the three message passing algorithms for a Poisson network with average degree $\lambda=4$ and $N=10^4$ nodes.

\begin{figure}[t]
	\includegraphics[width=1.0\columnwidth]{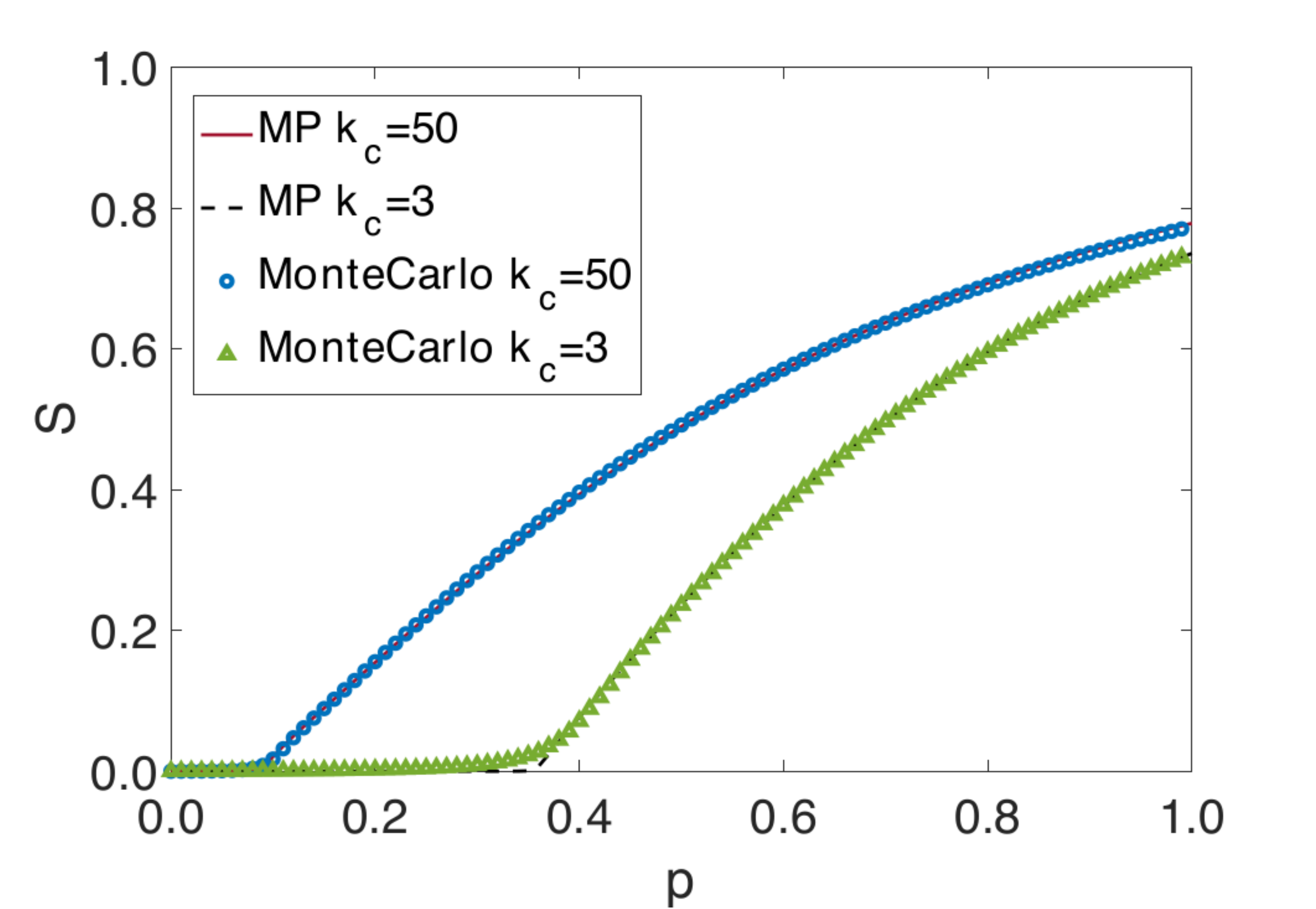}
	\caption{ (Color online)
	The fraction of infected nodes $S$ is plotted versus $p$ for  a uncorrelated scale-free network with $N=5\times 10^4$ nodes and power-law exponent $\gamma=2.5$. The results obtained by averaging the MonteCarlo simulations over over $200$ realization of the  configuration $\{T_i\}_{i\in V}$ and $\{x_{ij}\}_{(i,j)\in E)}$  are compared with the results of the MP algorithm defined by Eqs. (\ref{mes_av1b}) and Eq. (\ref{mes_av2b}). Here   $T(k)$ is given by Eq. (\ref{t_k_rhob}) with $\rho=0$,$\alpha=0$ and $k_c=50,3$.}\label{fig:0S}
\end{figure}

In the main text of this Letter we have shown the perfect agreement between the message passing algorithm define in Eq. (\ref{mes_av1b}) and Eq. (\ref{mes_av2b}) and the MonteCarlo simulations averaged over the distribution of $\{x_{i,j}\}_{(i,j)\in E}$ and the distribution of $\{T_i\}_{i\in V}$ in the case of a Poisson network. In Figure $\ref{fig:0S}$ we show that this excellent agreement also extend heterogeneous networks.

\begin{figure}[t]
	\includegraphics[width=\linewidth]{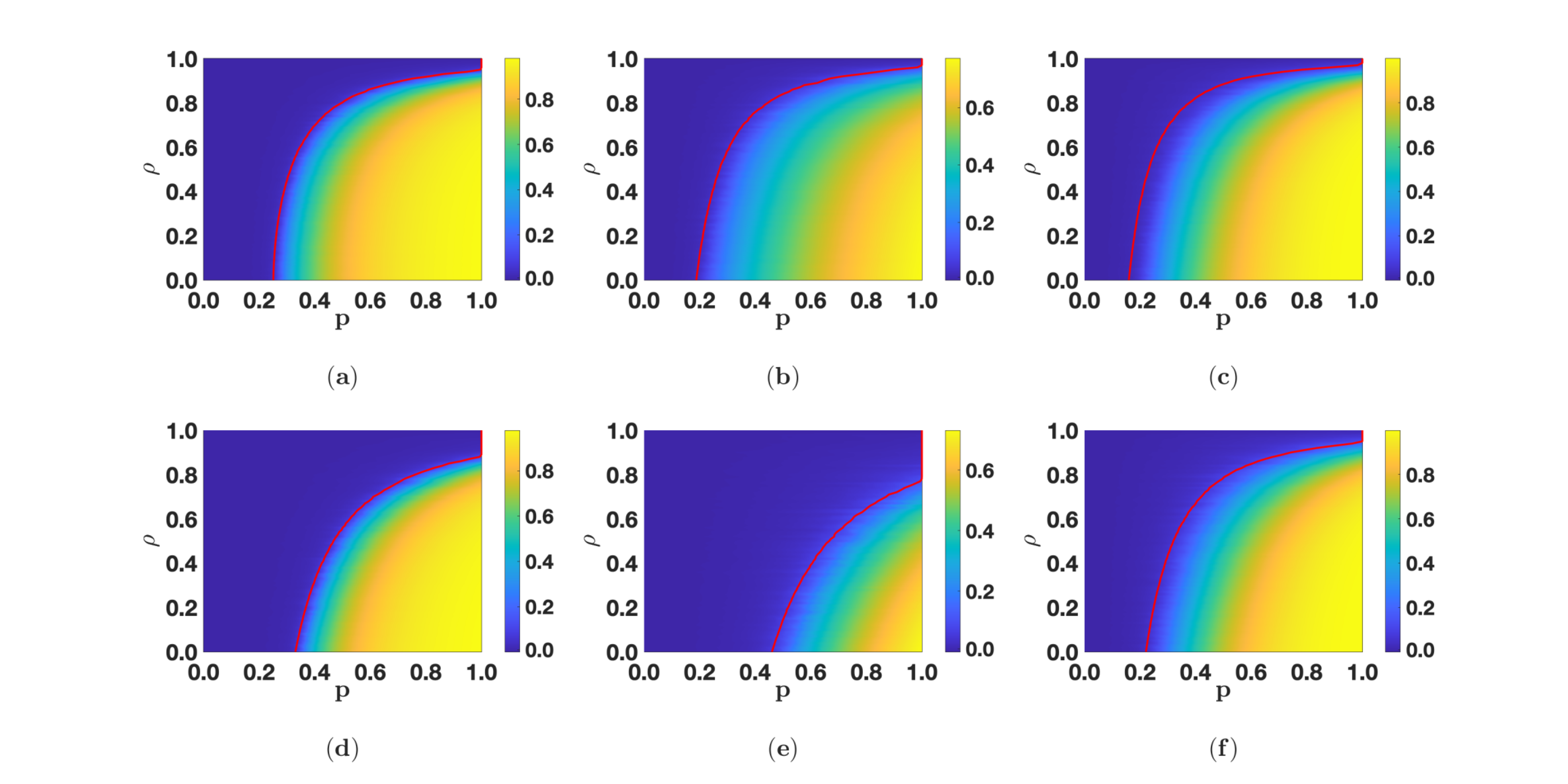}
	\caption{ (Color online) The phase diagram of the epidemic model is shown by plotting the fraction $S$ of infected individuals  obtained using the MonteCarlo algorithm in the plane $(p,\rho)$ for a  $N=10^4$-node networks and $T(k)$ given by Eq.(\ref{t_k_rhob}) with $\alpha=0$. The data are averaged $20$ times. The different panels correspond to different network topologies and different cutoffs $k_c$: Poisson network with $\lambda=4$ and $k_c=10$ (panel (a)) and $k_c=5$ (panel (d);  uncorrelated scale-free network with $\gamma=2.5$ and $k_c=10$ (panel (b)) and $k_c=3$ (panel (e)); BA network with $m=2$ and $k_c=10$ (panel (c)),  $k_c=5$ (panel (f)).The solid (red) lines indicate the epidemic threshold predicted by Eq. (\ref{pcb})}\label{fig:2S}
\end{figure}

We have also studied the results obtained averaging over several   MonteCarlo simulation for Poisson networks, BA network and for uncorrelated scale-free networks (see Figure $\ref{fig:2S}$) .
We found that the introduction of a non trivial cutoff $k_c$ can significantly increase the epidemic threshold $p_c$ well captured by Eq. (\ref{pcb}).

\section*{Comparison of different strategies of incrementing the adoption of the app on real datastes }

\begin{figure}[t]
	\includegraphics[width=\linewidth]{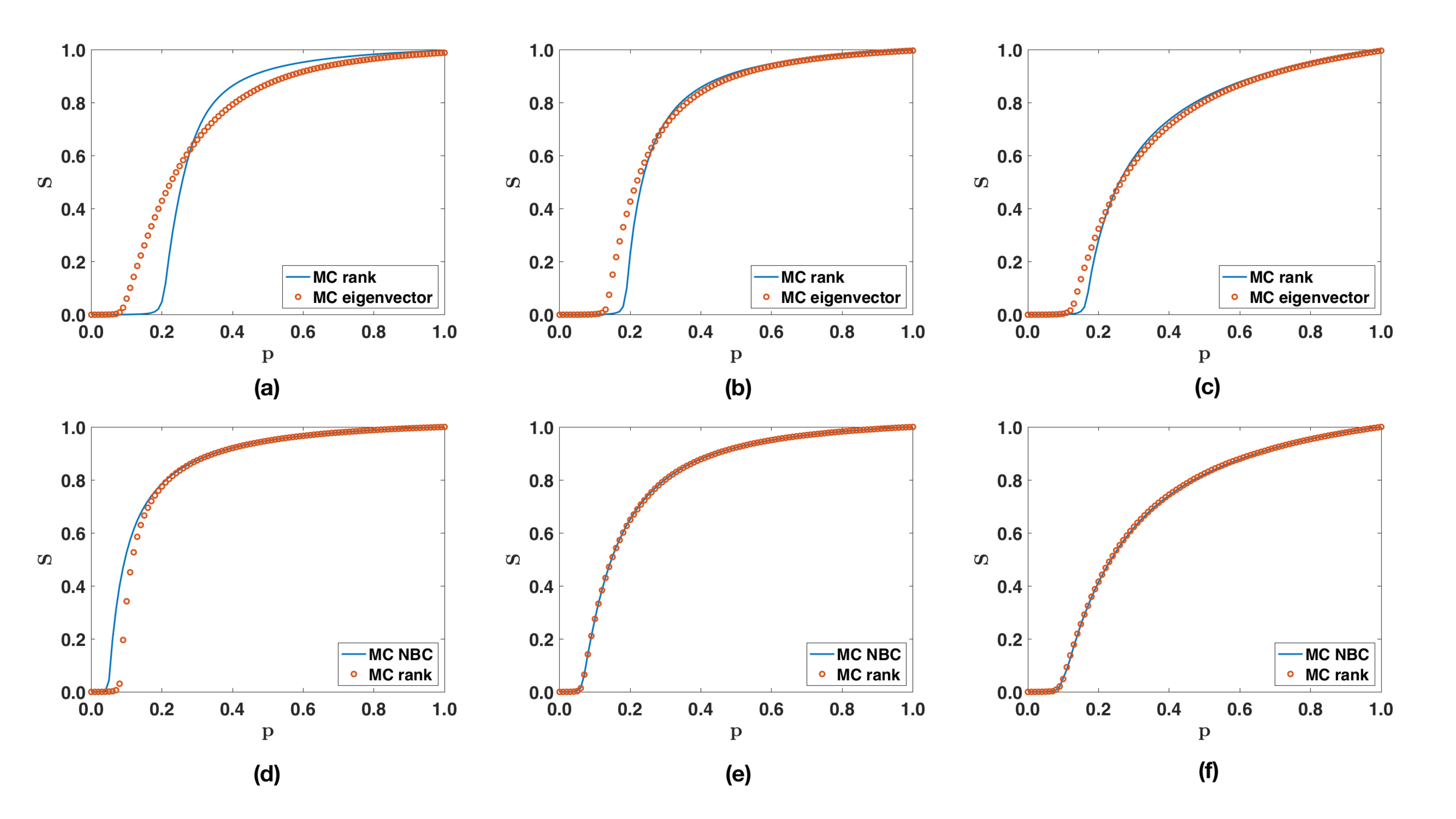}
	\caption{ We compare the efficiency of different strategies for targeting the adoption of the app including targeting the nodes of high degrees, the nodes of high eigenvector centrality and the nodes of high non-backtracking centrality (NBC) on three different real social network datasets.  Specifically we compare the fraction of infected individuals $S$  as a function of the transmissibility $p$ obtained with the MonteCarlo simulations  when the same  fraction $f$ of nodes of high centrality adopts the app but the centrality measures can change. In panels (a), (b) and (c) we report results obtained on the real datasets by comparing the strategy in which nodes of high eigenvector centrality are targeted  with the strategy in which the same fraction of high degree nodes are targeted. In panels (d) (e) and (f) a similar comparison is made between the strategy targeting nodes with high non-backtracking centrality and the strategy in which nodes of high degree are targeted. Panels (a)\&(d), (b)\&(e) and (c)\&(f) show the results obtained on  the friendship networks from the music streaming site Deezer in the countries of Croatia ($N = 54573$), Hungary ($N = 47538$) and Romania ($N = 41773$) respectively \cite{Rozemberczki_2019}. }
	\label{fig:centrality}
\end{figure}

In the main body of the paper we have shown that for  random uncorrelated networks when  the adoption of the app only depends on the degree of the nodes, targeting high degree nodes is the optimal strategy  for suppressing the epidemic spreading in a scenario of reduced resources. Here we want to investigate how this strategy compares to alternative strategies that target nodes with high eigenvector or high non-backtracking centrality \cite{rogers2015assessing} on real social network datasets.
In Figure $\ref{fig:centrality}$ we show the fraction of infected nodes $S$ versus $p$ when we assume that a fraction $f$ of highly central nodes adopt the app.  The centrality measures are taken to be the degree centrality, the eigenvector centrality, the non-backtracking centrality.
As  Figure $\ref{fig:centrality}$ shows, in the investigated datasets targeting nodes with high eigenvector centrality is not as efficient as targeting high degree nodes.  Targeting nodes with high non-backtracking matrix perform much better, however in the observed datasets  it does not appear to change significantly the results obtained by targeting the high degree nodes.
These numerical results suggest that in a wide-range of real scenarios, targeting high degree nodes can still be a very efficient algorithm for mitigating an epidemic outbreak.

\end{document}